# Recovering classical dynamics from coupled quantum systems through continuous measurement

Shohini Ghose,* Paul Alsing, and Ivan Deutsch

*Department of Physics and Astronomy, University of New Mexico, Albuquerque, New Mexico 87131*

Tanmoy Bhattacharya, Salman Habib, and Kurt Jacobs

*T-8 Theoretical Division, MS B285, Los Alamos National Laboratory, Los Alamos, New Mexico 87545*

We study the role of continuous measurement in the quantum to classical transition for a system with coupled internal (spin) and external (motional) degrees of freedom. Even when the measured motional degree of freedom can be treated classically, entanglement between spin and motion causes strong measurement back action on the quantum spin subsystem so that classical trajectories are not recovered in this mixed quantum-classical regime. The measurement can extract localized quantum trajectories that behave classically only when the internal action also becomes large relative to $\hbar$.

Quantum and classical mechanics offer differing predictions for the dynamics of a closed system specified by a given Hamiltonian. In recent years, it has been widely appreciated that emergent classical behavior can result when a quantum system is weakly coupled to an environment. Two levels of description have been used to discuss this behavior. The first utilizes the decoherence resulting from tracing over the environment to suppress quantum interference. In many circumstances this can lead to an effectively classical evolution of a phase-space distribution function [1]. A more fine-grained description is achieved when the environment is taken to be a meter that is continuously monitored, leading to a "quantum trajectory unraveling" of the system density operator conditioned on the measurement record [2]. If one averages over all possible measurement results, the description reverts to that at the level of phase-space distributions.

The quantum trajectory approach is a powerful tool for understanding and quantitatively identifying the quantum-classical boundary [3,4]. Continuous measurement provides information about the state of the system and thus localizes it in phase space. These localized trajectories have added quantum noise, however, due to quantum measurement back action. Therefore, in order to recover the desired classical trajectories, the system must be in a regime where the measurement causes strong localization but weak noise. The conditions for which both constraints are satisfied determine a system action scale for which classical dynamics can be observed in trajectories [4]. The trajectory approach provides a description of experiments [5] where a single quantum system is continuously monitored, thus enabling the study of quantum back action and the transition to classical dynamics in such systems. Continuous measurement records can also be employed in real-time feedback loops opening up an additional regime of quantum control [6].

Reference [4] dealt with measured systems with one motional degree of freedom. As the number of dynamical variables is enlarged, analyzing the quantum-classical transition becomes significantly more complex. Multiple coupled degrees of freedom can possess widely varying characteristic actions. Continuous measurement may be performed on any subset of the system variables with differing effects of quantum back action. Finally, the quantum state of the coupled system can become highly entangled so that the quantum back action can be nontrivially distributed among the various subsystems.

We present here the effect of continuous measurement on the dynamics of a particle with coupled spin and motional degrees of freedom. Numerical and analytical results demonstrate that, even if the position measurement satisfies the inequalities required for classical behavior, its entanglement with a quantum spin can still result in large measurement back action. Such a mixed quantum-classical description of systems with two coupled degrees of freedom is relevant in a variety of settings, including the Born-Oppenheimer description of molecules, polaron dynamics in condensed matter [7], and the transport of ultracold atoms in magneto-optic traps [8]. The last case is an experimentally clean system in which state preparation, manipulation, and measurement have been demonstrated [8,9]. The coupling between the motion of the atom in the lattice and its internal spin dynamics leads to entanglement at the quantum level and chaos in the classical description [10]. Although not restricted to this system, our analysis here is motivated by these experiments.

We take as our model Hamiltonian

$$H = \frac{p^2}{2m} + \frac{1}{2}m\omega z^2 + bzJ_z, \quad (1)$$

which describes a harmonic oscillator with mass $m$ and angular frequency $\omega$ with an additional spin- ($J_z$)-dependent constant force $bJ_z$, that can be interpreted as arising from a magnetic field with uniform gradient. This is none other than a Stern-Gerlach Hamiltonian, but with an additional trapping potential. The Hamiltonian also approximates a single lattice site of a one-dimensional (1D) "lin-angle-lin" optical lattice for trapping ultracold neutral atoms [8], where $z$ and $p$ are the center-of-mass position and the momentum of the atom. The spin $J_z$ is a constant of motion and determines the effective potential experienced by the center-of-mass motion: in the quantum case, eigenstates with $J_z = M_J$ experience a harmonic well centered at $(M_J/|J|)\Delta z$ with $\Delta z \equiv -b|J|/m\omega^2$.

---

*Electronic address: sghose1@unm.edu

The classical description generated by this Hamiltonian is that of a magnetic moment moving in a spatially inhomogeneous magnetic field while trapped in a harmonic well. Here we answer the question—*when does the continuous quantum measurement record follow the trajectory predicted by the classical Hamilton equations?* The measured signal can be viewed as a "mean" that is proportional to the expectation value of a quantum operator and a component that is well described as a white-noise process. An example of a measurement record is the photocurrent registered by the detector of a probe laser beam.

Assuming perfect measurement efficiency, the evolution of the system conditioned on a noisy record of the atom's position, $\langle z\rangle+(8k)^{-1/2}dW/dt$, may be studied using a stochastic Schrödinger equation [11],

$$d|\tilde\psi\rangle=\left\{\left(\frac{1}{i\hbar}H-kz^2\right)dt+(4k\langle z\rangle dt+\sqrt{2k}dW)z\right\}|\tilde\psi\rangle, \quad (2)$$

where $|\tilde\psi\rangle$ denotes an unnormalized quantum state, $k$ is the "measurement strength," and $dW$ describes a Wiener noise process. Rewriting Eq. (2) in terms of positive operator valued measures [12], we can evolve this equation numerically using a Milstein algorithm for the stochastic term. We pick as our initial condition a product of minimum-uncertainty coherent states in position and spin, and compare to the classical trajectories initialized with the same mean values of position, momentum, and spin direction; we choose the initial spin coherent state in the $x$ direction. We fix $b=-m\omega^2\Delta z/J$ with $\Delta z\approx 22 z_g$ where $z_g$ is the ground state root-mean-square (rms) width of the wells. We choose the action $I$, associated with the motion in the well, to be $I\approx 1000\hbar$ and a measurement strength $k=\omega/20z_g^2$ sufficient to observe classical dynamics of the positional degree of freedom uncoupled from the spin subsystem [4].

Consider first the behavior for the smallest spin system, $J=1/2$. We see in Fig. 1(a) that the quantum trajectory quickly diverges from the classical trajectory. This can be understood by noting that the initial quantum state pointing in the $x$ direction is an equal superposition of spin-up and spin-down states that move along the wells centered at $z_\uparrow=-\Delta z$ and $z_\downarrow=\Delta z$, respectively [Figs. 2(a) and 2(b)]. The two spin components of the initial spatially localized wave packet thus separate into a left and a right wave packet, so that the total wave function evolves into an entangled Bell-like state, $|\psi(t)\rangle=|\phi_L\rangle|\uparrow\rangle+|\phi_R\rangle|\downarrow\rangle$, with $\langle\phi_L|\phi_R\rangle$ rapidly decreasing from unity. As the two components $\phi_L$ and $\phi_R$ experience different potentials, this splitting of the wave packet is reflected in an initial rapid increase of the variance in position of the wave function [outer solid curves in Fig. 1(a)]. Eventually, these components become resolvable beyond measurement errors, and the measurement collapses the wave function into one of the wells. In contrast, the classical dynamics predicts that the spin precesses freely and the particle experiences the average of the "left" and "right" potentials.

Because the Hamiltonian under consideration is *linear*, the distinct behavior of the quantum and classical dynamics arises *solely* from the choice of initial condition. Indeed, if we choose an initial statistical mixture with equal probability to have the magnetic moment aligned and antialigned along $z$, the classical evolution produces the same probability distribution of the $z$ component of the spin as that associated with the spin-1/2 quantum particle initially polarized along $x$. For an initial coherent state, however, no classical distribution can match all observables. The particular choice of ini-tial condition described above matches the distribution of $J_z$ at the expense of the other components of the spin. If the system is generalized by adding a transverse magnetic field, so that the $x$ and $y$ components affect the dynamics, these initial conditions will also fail to reproduce the quantum behavior.

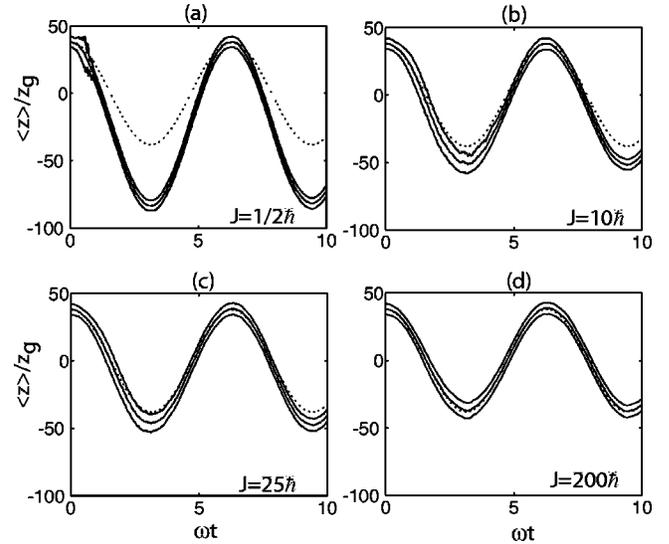

FIG. 1. Mean position of the measured system (solid) for different values of spin with $\Delta z\approx 22 z_g$, $I\approx 1000\hbar$, $k=\omega/20z_g^2$. Outer solid curves show the variance of the wave function. As $J$ gets larger the mean position approaches the classical (dotted) trajectory.

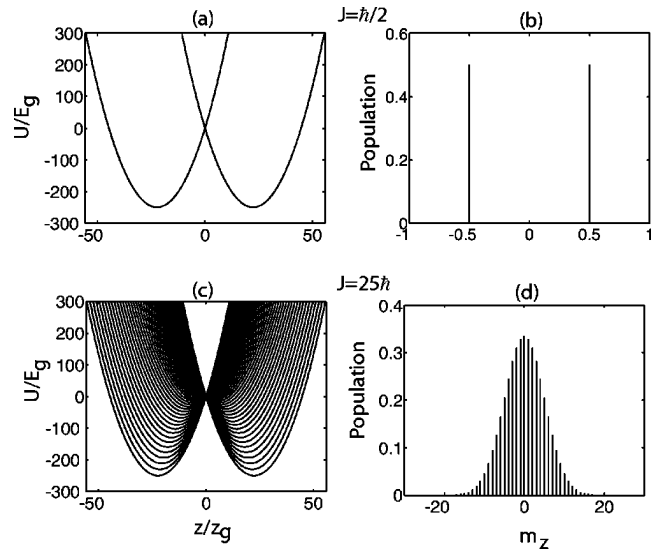

FIG. 2. The spin-up and spin-down components of a spin-1/2 wave function move along two different potential wells (a), (b). For $J\gg 1/2$, the spin components of the wave function evolve along $2J+1$ different potentials (c). Histograms for the populations in each $m_z$ state for a spin-coherent state in the $x$ direction (d) show that as $J$ gets larger the population becomes peaked around the $m_z=0$ state. The position is thus more likely to localize the wave function in the central (classical) potential well.

With these factors in mind, in this study, we have generically chosen to match the initial mean values of position, momentum, and spin direction. The spin-coherent state in the $x$ direction, having a nonzero variance in $J_z$, differs dramatically from the classical state with mean moment along $x$. This difference affects the evolution of the mean values in the measurement record in a profound way, as we saw above.

Unlike the spin-1/2 case, for a large $J$, an initial spin-coherent state in the $x$ direction is no longer a superposition solely of spin-up and spin-down states in the $J_z$ basis, but rather a distribution over all $2J+1$ $M_J$ states, peaked at $M_J=0$. Just as in the spin-1/2 case, an initially localized wave function will spread out in space as the different spinor components move along the different potentials centered at $z_{M_J} = -(M_J/J)\Delta z$ [Figs. 2(c) and 2(d)]. However, as $J$ becomes larger, the population distribution becomes more peaked at the $M_J=0$ state and the potentials experienced by the different components are more similar. Most of the population moves along potentials centered near $z_0=0$, which are close to the classical potential. The measurement is thus more likely to localize the atom around the classical potential and damp out the tails of the wave function that spread out over the outermost potentials. The key point is that position measurement no longer results in a strongly projective spin measurement, and therefore the weak-noise condition can be met along with the strong localization condition.

We can determine analytically the scale of $J$ for which the weak-noise and strong-localization conditions are satisfied by generalizing the approach in Ref. [4]. The stochastic equations of motion for the mean position and momentum follow from Eq. (2),

$$d\langle z \rangle = \frac{\langle p \rangle}{m} dt + \sqrt{8k} C_{zz} dW,$$

$$d\langle p \rangle = -m\omega^2 \langle z \rangle dt - b \langle J_z \rangle dt + \sqrt{8k} C_{zp} dW, \quad (3)$$

$$d\langle J_z \rangle = \sqrt{8k} C_{zJ_z} dW,$$

where $C_{ab} = (\langle ab \rangle + \langle ba \rangle)/2 - \langle a \rangle \langle b \rangle$ are the symmetrized covariances. We have not included here the $x$ and $y$ components of the angular momentum since the position and momentum equations depend only on $J_z$. Because our system is linear, these equations are the same as those for a classical stochastic process. Nevertheless, because in our *initial condition* the variances and higher moments of the quantum state do not agree with the classical ones, the quantum expectation values will not, in general, follow the classical trajectories. A measurement will be a faithful record of the classical equation only when the covariance matrix elements remain small at all times relative to the allowed phase space of the dynamics.

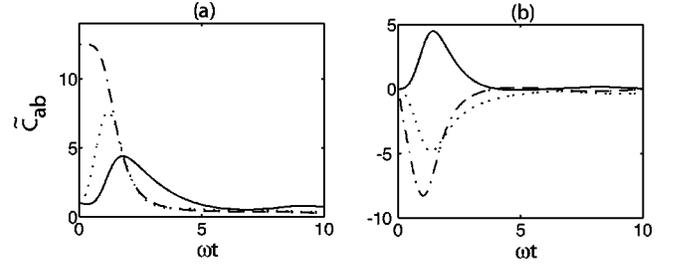

FIG. 3. Solutions of $C(t)$ for $J=25\hbar$ and $\Delta z$, $I$ and $k$ chosen to be the same as in Fig. 1: (a) variances $C_{zz}/z_g^2$ (solid), $C_{pp}/p_g^2$ (dotted), and $C_{J_z J_z}/\hbar^2$ (dash-dotted); and (b) covariances $C_{zp}/z_g p_g$ (solid), $C_{zJ_z}/\hbar z_g$ (dotted), and $C_{pJ_z}/\hbar p_g$ (dash-dotted).

To determine the evolution of the cumulants, we truncate the hierarchy of coupled cumulant equations, and neglect third and higher cumulants since these remain small as $J$ becomes large. Even under this approximation, although analytical solutions for the second cumulants $C(t)$ can be found as solutions to a matrix Riccati equation [13], they are in general not simple functions of the system parameters. Numerical studies show that the upper bounds on the magnitude of $C(t)$ decrease as $J$ is increased, as expected. Figure 3 shows a typical plot of $C(t)$ with $J=25\hbar$ and $I\approx 1000\hbar$. For this value of $J$, the maximum variance in the measured position is already smaller than the allowed phase space of the dynamics [Fig. 1(c)] by a factor of about 50.

Our numerical and analytical results show that classical dynamics is recovered in this coupled system only when the actions of both subsystems become large relative to $\hbar$. When one subsystem lies in the quantum regime, even a weak measurement of the classical subsystem eventually results in a strongly projective measurement of the quantum subsystem, thus preventing the recovery of classical behavior. The dynamics of such coupled systems has previously been approximated in other contexts using a mixed quantum-classical description [14], where the classical subsystem variables are described by $c$ numbers while the quantum subsystem variables are treated as quantum mechanical operators. In particular, the work by Schanz and Esser [14] treats exactly the same model Hamilton as we consider here, but with the addition of a transverse magnetic field. Such a description results in equations of motion for the means that can lead to chaotic dynamics. However, in the absence of an appropriate separation of time scales, the prediction of chaos arising from such a mixed quantum-classical description has been shown to fail [15]. Here we find that even in a regular *nonchaotic regime*, for open systems, this mixed quantum-classical description will have limited validity since it does not take into account the effect of measurement back action when the system is actually observed. While the Hamilton and Heisenberg equations of motion will certainly agree in the ensemble mean, variances will be quite large so that the measured individual quantum trajectories will diverge strongly from the mean. The same argument should hold for open systems in the chaotic regime.

Beyond the behavior of the observables of each marginal subsystem (internal and external), entanglement in the whole system characterizes the quantum to classical transition. In the spin-1/2 case we have seen that the state can develop nearly maximal entanglement, when the spin-up and spin-down wave packets become spatially resolved. For larger values of $J$, the evolution never results in such resolution, and hence the entanglement, which is determined by the overlap between the different spinor components of the wave function, also decreases as $J$ increases. For pure bipartite systems, a measure of the entanglement is given by the von Neumann entropy of the marginal density matrix $\tilde{\rho}$ for either subsystem, $E = -\text{Tr}(\tilde{\rho} \ln \tilde{\rho})$. The degree of entanglement generated in stochastic dynamics can be compared for different values of $J$ by calculating the normalized value $\tilde{E}_{\max} = E_{\max}/E_0$, where $E_{\max}$ is the maximum entropy achieved during the monitored evolution and $E_0$ is the maximum possible entropy for the chosen initial state. We find that $\tilde{E}_{\max}$ falls off with increasing $J$ as expected, and with a $1/\sqrt{J}$ dependence for large $J$. These results raise intriguing questions about the role of entanglement between subsystems in the transition to classical behavior.

In future work we hope to investigate these issues further. Our model Hamiltonian becomes nonintegrable when an additional magnetic field in the $x$ direction is applied. We propose to generalize our results to this nonintegrable regime and study the emergence of classical chaos through continuous measurement and decoherence.

We thank Poul Jessen and Daniel Steck for helpful discussions. S.G., P.M.A., and I.H.D. were supported under NSF Grant No. PHY-009569.

[1] S. Habib, K. Shizume, and W. H. Zurek, Phys. Rev. Lett. **80**, 4361 (1998).

[2] H. Carmichael, *An Open Systems Approach to Quantum Optics* (Springer-Verlag, Berlin, 1993).

[3] T. P. Spiller and J. F. Ralph, Phys. Lett. A **194**, 235 (1994); M. Schlautmann and R. Graham, Phys. Rev. E **52**, 340 (1995); T. Brun, I. Percival, and R. Schack, J. Phys. A **29**, 2077 (1996).

[4] T. Bhattacharya, S. Habib, and K. Jacobs, Phys. Rev. Lett. **85**, 4852 (2000).

[5] H. Mabuchi, J. Ye, and H. J. Kimble, Appl. Phys. B: Lasers Opt. **68**, 1095 (1999); P. Warszawski, H. M. Wiseman, and H. Mabuchi, Phys. Rev. A **65**, 023802 (2001); S. A. Gurvitz, L. Fedichkin, D. Mozyrsky, and G. P. Berman, e-print cond-mat/0301409.

[6] H. M. Wiseman, Phys. Rev. A **49**, 2133 (1994); A. C. Doherty, S. Habib, K. Jacobs, H. Mabuchi, and S. M. Tan, *ibid.* **62**, 012105 (2000).

[7] T. Holstein, Ann. Phys. (N.Y.) **8**, 325 (1959); D. Hennig and B. Esser, Phys. Rev. A **46**, 4569 (1992).

[8] I. H. Deutsch and P. S. Jessen, Phys. Rev. A **57**, 1972 (1998).

[9] D. L. Haycock, P. M. Alsing, I. H. Deutsch, J. Grondalski, and P. S. Jessen, Phys. Rev. Lett. **85**, 3365 (2000).

[10] I. H. Deutsch, P. M. Alsing, J. Grondalski, S. Ghose, D. L. Haycock, and P. S. Jessen, J. Opt. B: Quantum Semiclassical Opt. **2**, 633 (2000); S. Ghose, P. M. Alsing, and I. H. Deutsch, Phys. Rev. E **64**, 056119 (2001).

[11] V. P. Belavkin and P. Staszewski, Phys. Lett. A **140**, 359 (1989); G. J. Milburn, K. Jacobs, and D. F. Walls, Phys. Rev. A **50**, 5256 (1994); A. C. Doherty and K. Jacobs, *ibid.* **60**, 2700 (1999); A. C. Doherty, K. Jacobs, and G. Jungman, *ibid.* **63**, 062306 (2001); C. W. Gardiner, *Handbook of Stochastic Methods* (Springer-Verlag, Berlin, 2002).

[12] C. M. Caves and G. J. Milburn, Phys. Rev. A **36**, 5543 (1987).

[13] W. T. Reid, *Riccati Differential Equations* (Academic, New York, 1972).

[14] P. Belobrov, G. Zaslavskii, and G. K. Tartakovskii, Sov. Phys. JETP **44**, 945 (1976); P. W. Milonni, J. R. Ackerhalt, and H. W. Galbraith, Phys. Rev. Lett. **50**, 966 (1983); R. Blumel and B. Esser, *ibid.* **72**, 3658 (1994); H. Schanz and B. Esser, Phys. Rev. A **55**, 3375 (1997).

[15] F. Cooper, J. Dawson, S. Habib, and R. D. Ryne, Phys. Rev. E **57**, 1489 (1998); L. E. Ballentine, *ibid.* **63**, 056204 (2001).